\documentclass[pra,aps,amssymb,twocolumn,superscriptaddress,
showpacs,showkeys,preprintnumbers,nofootinbib,floatfix]{revtex4}
 
\usepackage{graphicx}
\usepackage{natbib}
\usepackage{latexsym}
\usepackage{amsfonts}
\usepackage{url,hyperref}
\usepackage{bm}
\usepackage{color}
\newcommand{\bb}{\mathbf}
\newcommand{\nn}{\nonumber\\}
\newcommand{\beq}{\begin{equation}}
\newcommand{\eeq}{\end{equation}}
\newcommand{\bed}{\begin{displaymath}}
\newcommand{\eed}{\end{displaymath}}
\def\bea{\begin{eqnarray}}
\def\eea{\end{eqnarray}}
\newcommand{\veps}{\varepsilon}
\newcommand{\unit}{\hat\mathbf}
\newcommand{\bnab}{\bm\nabla}

\begin{document}

\title{Dynamical Casimir Effect for TE and TM Modes in a \\ Resonant Cavity 
Bisected by a Plasma Sheet} 
\author{W. Naylor}
\altaffiliation{Author for correspondence: naylor@se.ritsumei.ac.jp}
\affiliation{Department of Physics, Ritsumeikan University, 
Kusatsu, Shiga, 525-8577, Japan}
\author{S. Matsuki}
\affiliation{Department of Physics, Ritsumeikan University, 
Kusatsu, Shiga, 525-8577, Japan}
\author{T. Nishimura}
\affiliation{Research Center of Ion Beam Technology, Hosei University, 
Koganei, Tokyo, 184-8584, Japan}
\author{Y. Kido}
\affiliation{Department of Physics, Ritsumeikan University, 
Kusatsu, Shiga, 525-8577, Japan}
\date{\today}

\begin{abstract}
Parametric photon creation via the dynamical Casimir effect (DCE) is evaluated numerically, in a three-dimensional rectangular resonant cavity bisected by a semiconductor diaphragm (SD), which is irradiated by a pulsed laser with frequency of GHz order. The aim of this paper is to determine some of the optimum conditions required to detect DCE photons relevant to a novel experimental detection system. We expand upon the thin plasma sheet model [Crocce et al., Phys. Rev. A {\bf 70} 033811 (2004)] to estimate the number of photons for both TE and TM modes at any given SD position. Numerical calculations are performed considering up to 51 inter-mode couplings by varying the SD location, driving period and laser power without any perturbations.  It is found that the number of photons created for TE modes strongly depends on SD position, where the strongest enhancement occurs at the midpoint (not near the cavity wall); while TM modes have weak dependence on SD position. Another important finding is the fact that significant photon production for TM$_{111}$ modes still takes place at the midpoint even for a low laser power of $0.01~\mu$J/pulse, although the number of TE$_{111}$ photons decreases almost proportionately with laser power.  We also find a relatively wide tuning range for both TE and TM modes that is correlated with the frequency variation of the instantaneous mode functions caused by the interaction between the cavity photons and conduction electrons in the SD excited by a pulsed laser.
\end{abstract}


\maketitle

 \section{Introduction} 
 
\par Motion induced radiation, or the dynamical Casimir effect (DCE) as it is now more commonly known, was first discussed by Moore \cite{moore:2679} in 1970 who showed that photons would be created in a Fabry-P\'erot cavity if one of the ends of the cavity wall moved with periodic motion. The mechanical oscillation amplitude of the wall is small: $v/c\ll 1$, where $v$ is the wall velocity and $c$ is the speed of light and in this limit the number of photons produced during a given number of parametric oscillations is proportional to $\sinh^2 (2\omega t \,v/c)$,  which was first discussed in \cite{Narozhnyi, Mostepanenko} (for a nice review see, e.g. \cite{Kim:2006zzj}). A  realistic cavity has a finite quality factor, $Q$, and any exponential growth eventually saturates proportionally to $\sinh^2 (2Q\, v/c)$, for times greater than the characteristic timescale $\tau=Q/ \omega$, e.g., see \cite{Kim:2006zzj}. The emission of radiation from  accelerated charges is a well known classical effect in electrodynamics. However, what is unusual with the DCE is the prediction that radiation can also be emitted from neutral moving objects. 

\par Aside from mechanical harmonic oscillations of a (cavity) wall \cite{BrownHayes:2006ip,Kim:2006zzj}, it is also possible to induce temporal variations of the dielectric function \cite{PhysRevA.47.4422,PhysRevA.49.433,PhysRevA.51.4109,JJAP.34.4508,JPSJ.65.3513,Antunes:2003jr,PhysRevLett.93.193601}, also see \cite{Mendonca:2008bg}. This leads to an {\it effective} wall motion by, for example, varying the optical path length of the cavity \cite{PhysRevA.51.4109,JPSJ.65.3513,Mendonca:2008bg}. A related approach to {\it effective} wall motion is to irradiate a semiconductor sheet with a pulsed laser, which leads to a plasma mirror with time dependent surface conductivity, e.g., see \cite{Yablonovitch:1989zza}. (Note that pulsed lasers also have applications in controlling the static Casimir force \cite{chen:035338}.) This has been modeled by using thin dielectric slabs \cite{PhysRevLett.93.193601, Dodonov:2005ob, Dodonov:2006is,Dodonov:2006jb}, or by using thin conducting slabs \cite{Crocce:2004jq}. Indeed experiments are already being built to detect DCE photons, based on the idea of using a semiconductor sheet irradiated by a pulsed laser \cite{Braggio:2005epl,Agnesi:2008ja} in centimeter sized cavities (for ideas relating to microcavities see e.g., \cite{liberato:103602,Gunter:07838}.) In order to verify the DCE experimentally, it is important to consider some of the optimum conditions, which guides one to design and set up a detection system. 

\par In this paper we extend the model proposed by Crocce et al. \cite{Crocce:2004jq} to include both TE and TM modes for a three-dimensional resonant cavity ($L_x\times L_y\times L_z$) bisected by a semiconductor diaphragm (SD).  The SD is irradiated by a pulsed laser, which provides a periodic change between semiconducting and metallic states.  Our primary concern is the dependence of the  number of created photons upon {the}  location of the SD, which can be performed  only by numerical calculations, because perturbative treatments can no longer be applied to a general SD position away from the cavity wall at low temperatures.  From an experimental point of view, it is desirable to divide the cavity into two parts, (i) a photon-creation/detection chamber and (ii) a pulsed-laser irradiation space to avoid the background caused by laser irradiation.  Another important factor to verify the DCE would be  to keep the cavity at a low temperature of $\sim100$ mK to suppress the number of thermal blackbody photons to less than unity.  Thus, numerical calculations are also performed to evaluate the number of created photons under the conditions of high ($50 \mu$J/pulse) and sufficiently low ($0.01\mu$J/pulse) laser powers.  The results obtained are discussed in connection with our proposed DCE detection system (see Sec.~\ref{assump}) using highly-excited Rydberg atom beams (Rb or K), which was already  successfully applied to explore the dark matter axion \cite{Bradley:2003kg}.

\section{Formulation}
\label{form}

\par One of the primary concerns of this work is the SD-position dependent enhancement of DCE photon creation. Thus, we have extended the model of Crocce et al. \cite{Crocce:2004jq}, see Appendix~\ref{AppA}, to both TE and TM modes for any SD position, $\eta=d/L_z$, where $d$ is the distance of the SD from the cavity wall and $L_z$ is the longitudinal length of the cavity. 
Their work \cite{Crocce:2004jq} is closely related to the 1D plasma sheet model of Barton \& Calogeracos \cite{Barton:1995iw}. In Appendix~\ref{AppA} we give the 3D generalization, where Maxwell's equations are separated using scalar Hertz potentials \cite{Nisbet:1955, Bordag:2005qv, Crocce:2005htz}, leading to two Klein-Gordon like equations, given below.
\par For TE modes a thin plasma sheet irradiated by a pulsed laser satisfies the following {\it jump} conditions (see Appendix \ref{AppA}) {at SD location $z=d$}:
\bea
\label{TEbc}
{\rm disc} \,\Psi(\bb r_{\bot},d,t) &=&0, \nn
{\rm disc} \,\Psi'(\bb r_{\bot},d,t) &=& {e^2 n_s(t)\over m^*}\,\Psi(\bb r_{\bot},d,t)~,
\eea
where $e$ is the electron charge, $n_s(t)$ is the surface number density of electrons excited to the conduction band, and $m^*$ is the effective mass of the conduction electrons {(${\bb r}_\bot$ corresponds to transverse directions)}. Note that ${\it disc}\,\Psi(d) \equiv \Psi(d_+)-\Psi(d_-)$. It is easy to show that the above {\it jump} conditions can be  derived from the following self-adjoint differential equation
\beq
 \bnab_{\bot}^2\Psi(\bb r,t) + \Psi''(\bb r,t)-\ddot\Psi(\bb r,t)={e^2 n_s(t)\over m^*} \delta (z-d)\Psi(\bb r,t)~,
 \label{ode}
\eeq
where $\Psi''$ and $\ddot \Psi$ represent derivatives with respect to $z$ and $t$, respectively. The orthonormalized mode functions for a thin SD plasma sheet located at $z=d$, then read
\begin{widetext}
\beq
\Psi_{\bf m} = 
\left\{
\begin{array}{ccc}
A^{\rm (TE)}_m\sqrt{\frac{1}{d}}\sin\,(k_{m_z} z)~v_{\bf k_\bot}(\bf r_{\bot}) ~,  & \qquad  \qquad 0<z<d \\
 B^{\rm (TE)}_m\sqrt{\frac{1}{L_z-d}}\sin\,(k_{m_z}(L_z-z))~v_{\bf k_\bot}(\bf r_{\bot})~,  &\qquad  \qquad d<z<L_z
\end{array}
\right.
\eeq
\end{widetext}
where the TE transverse mode functions, $v_{\bf k_\bot}$, are
\beq
v_{\bf k_\bot}({\bf r}_{\bot}) = 
\sqrt{\frac{2}{L_x}}\cos\left(\frac{\pi m_x  x}{L_x}\right)\sqrt{\frac{2}{L_y}}
\cos\left(\frac{\pi m_y  y}{L_y}\right),
\eeq
with
$\bnab_{\bot}^2 v_{\bf k_\bot} = -{\bf k}_\bot^2 v_{\bf k_\bot}$. The case for TM modes is a little more delicate, but a short calculation leads to (see Appendix \ref{AppA}) the {\it jump} conditions:
\bea
\label{TMbc}
{\rm disc} \,\Phi(\bb r_\bot,d,t) &=&{e^2 n_s(t)\over {\bf k}_\bot^2 m^*}
\Phi'(\bb r_\bot,d,t), \nn
{\rm disc} \,\Phi'(\bb r_\bot,d,t)& =&0~.
\eea
These {\it jump} conditions can also be derived from the following wave equation, 
\bea
\bnab_{\bot}^2\Phi(\bb r,t) +  \Phi''(\bb r,t)-\ddot\Phi(\bb r,t)&=&{e^2 n_s(t)\over {\bf k}_\bot^2 m^*}\delta'(z-d)\Phi(\bb r,t)~;\nn
\label{TMode}
\eea
however, this representation is not, strictly speaking, mathematically correct, e.g., see \cite{Solva}. The orthonormalized mode functions for TM modes are
\begin{widetext}
\beq
\Phi_{\bf m} = 
\left\{
\begin{array}{ccc}
A^{\rm (TM)}_m\sqrt{\frac{1}{d}}\cos\,(k_{m_z} z)~u_{\bm k_\bot}(\bf r_{\bot})~, &
\qquad \qquad 0<z<d \\
 B^{\rm (TM)}_m\sqrt{\frac{1}{L_z-d}}\cos\,(k_{m_z}(L_z-z))~u_{\bf k_\bot}(\bf r_{\bot}) ~,  & \qquad\qquad d<z<L_z
\end{array}
\right. 
\eeq
\end{widetext}
where the TM transverse mode functions are
\beq
u_{\bm k_\bot}({\bf r_{\bot}}) =
\sqrt{\frac{2}{L_x}}\sin\left(\frac{\pi m_x  x}{L_x}\right)\sqrt{\frac{2}{L_y}}
\sin\left(\frac{\pi m_y  y}{L_y}\right)
\label{TMperp}
\eeq
with 
$\bnab_{\bot}^2 u_{\bf k_\bot} = -{\bf k}_\bot^2 u_{\bf k_\bot}$. Finally, by imposing continuity  in conjunction with the jump conditions,  
we find the following eigenvalue relation:                                                                                                                                                                                      
\beq
{ \sin(k_{m_z} L_z)\over (k_{m_z})^{\mp 1} \sin(k_{m_z}[L_z-d]) \sin(k_{m_z} d)} =  \mp{1\over {\bf k}_\bot^2} {e^2 n_s(t)\over m^*}~,
\label{transTETM}
\eeq
where the $\mp$ signs refer to TE and TM modes respectively (for TE drop the factor $1/{\bf k}^2_\bot$). In the following we shall work with the dimensionless variable $V(t) L_z= (e^2 n_s(t) /m^* )L_z$ and thus, for TM modes the potential scales as $ V(t) L_z/({\bf k}_\bot^2 L_z^2 )$. The time variation of $k_{m_z}(t)$ leads to a frequency variation in the angular frequency, $\omega_{\bb m}(t)$, see next section.

\section{Quantization}
\label{quant}

\par In the plasma sheet model there is no time dependence in the bulk dielectric permittivity and so there are no issues with quantization in time-dependent dielectrics, e.g., see discussion in \cite{PhysRevLett.93.193601}. Thus, quantization can be made straightforwardly with the quantum field operator expansion of the Hertz scalars \cite{Hacyan:1990ja}:
\beq
\widehat{\psi}({\bf r},t) =\sum_{\bf m} 
\left[ a_{\bf m} \psi_{\bf m}({\bf r}, t) + a^\dagger_{\bf m} \psi^*_{\bf m}({\bf r}, t) \right]~,
\eeq
for TE modes, while expressions
for TM modes are made by identifying $\psi \to\phi$. 
The normalization is set to unity for reasons explained below. 
To evaluate the number of photons produced in a given mode we shall use 
the Bogolubov method by first defining, for $t\geq 0$ (during irradiation) an {\it instantaneous} basis \cite{Law:1995zz}: 
\beq
 \psi^{\rm out}_{\bf s}(\mathbf r, t) = \sum_{\bf m} P_{\bf m}^{({\bf s})} \Psi_m(\bb r, t)~,
 \label{out}
 \eeq 
 with a similar expression for the TM component in terms of $\Phi$. For  TE modes with $t\leq 0$ (before irradiation) we have
\beq
\psi^{\rm in}_{\bf m}(\bb r, t) = {{\rm e}^{-i\omega^0_m t}\over\sqrt{2\omega_m^0}}
\sqrt{\frac{2}{L_z}}\sin\left(\pi m_z z\over L_z\right)~v_{\bf k_\bot}(\bf r_{\bot}) 
\label{in}
\eeq
(for TM modes replace $\sin \to \cos$ and $v_{\bf k_\bot}\to u_{\bf k_\bot}$) 
where the stationary angular frequency is
\beq
\omega_{\bf m}^0=\omega_{\bf m}(0)=  c \pi \sqrt{\left(\frac{m_x}{L_x}\right)^2 + \left(\frac{m_y }{L_y} \right)^2 +\left(\frac{m_z }{L_z} \right)^2} ~,
\label{statfreq}
\eeq 
where $c$ is the speed of light in vacuum.
During irradiation ($t\geq 0$) the angular frequency becomes time dependent:
\beq
\omega_{\bf m}^2(t) = c^2 \Big[ \left(\frac{m_x
\pi}{L_x}\right)^2 + \left(\frac{m_y \pi}{L_y} \right)^2+k_{m_z}^2(t) \Big]~,
\label{omega}
\eeq
where $k_{m_z}(t)$ is the eigenvalue given in Eq. (\ref{transTETM}). 
Substitution into the wave equation, (\ref{ode}) or (\ref{TMode}), on either side of the SD then leads to \cite{Law:1995zz, Schutzhold:1997yh,Crocce:2004jq}
\bea
\ddot P_n^{(s)} +  \omega_n^2 (t) P_n^{(s)} &=& - \sum_m^{\infty} \left[ 2 M_{m n} \dot P_m^{(s)}+\dot M_{m n} P_m^{(s)} \right.\nn
&&+ \left.   \sum_\ell^\infty M_{n\ell} M_{m\ell } P^{(s)}_m \right]~,
\label{Osc}
\eea
where the terms $\omega_m(t)$ and 
\beq
M_{ m n}= \left(\Psi_n, \Psi_n
\right)^{-1}\delta_{m_x n_x}\delta_{m_y n_y}
\,   \left(  \frac{\partial \Psi_m
}{\partial t} , \Psi_n \right)~,
\label{emm}
\eeq  correspond to squeezing and acceleration terms respectively \cite{Schutzhold:1997yh}. It is important to note that because of symmetry, the relation for  $M_{\bf m n}$ in the transverse ($x,y$) directions reduces to simple Kronecker deltas and hence, the mode functions reduce to an effective 1D problem \cite{Ruser:2005xg}; the vector indices also become one-dimensional: ${\bf m} (m_x,m_y,m_z)\to m_z=m$. This is also the reason why the normalisation is set to unity. As opposed to a vibrating cavity wall, simple sinusoidal expressions for $M_{mn}$ do not exist, although for the plasma sheet  model complicated analytic expressions can be found  for a given $\eta$.
  
By defining auxiliary functions the problem reduces to a system of coupled first-order differential equations \cite{Ruser:2005xg,Razavy:1985un}, also see \cite{JPSJ.65.3513}. These are solved by truncating the infinite sums with a cut-off in the mode sums such that $\ell\to \ell_{max}$ does not change the result in Eq. (\ref{Nmode}) below, as we have verified below.\footnote{The number of created photons saturates at a cut-off of $\ell_{max}\gtrsim 50$ in the  mode sums.} 
The Bogolubov coefficients are defined as \cite{Birrell} 
 \beq
 \alpha_{mn}=(\psi_m^{\rm out},\psi_n^{\rm in})\,,\quad
 \quad
 \beta_{mn}=-(\psi_m^{\rm out},[\psi_n^{\rm in }]^*)
 \label{bog}
 \eeq
 where the invariant scalar product is 
 $(\phi,\psi)=-i\int_{\rm cavity} d^3 x (\phi\,\dot \psi^* - \dot \phi\, \psi^*)$. Then, in terms of the original  functions in Eqs. (\ref{out}) and (\ref{in}), for {\it out} and {\it in} modes respectively, we find
\beq
 \beta_{mn}=\sqrt{\omega_m\over2}P_m^{(n)}
-i\sqrt{1\over2\omega_m}\Big[\dot{P}_m^{(n)}+
\sum_\ell^{\ell_{max}} M_{\ell m} P_\ell^{(n)}\Big] \,,
\label{Pees}
\eeq
where  $ \alpha_{mn}$ is obtained by complex conjugation. The number of photons in a given mode (assuming an initial vacuum state) is given by\footnote{The number of photons created requires  regularization such as by introducing an explicit frequency cut-off \cite{Schutzhold:1997yh}.} 
\beq
N_m(t) = \sum_n^{\ell_{max}} |\beta_{mn}|^2~,
\label{Nmode}
\eeq 
where the total number of photons (if needed) is given by
\beq
N(t) = \sum_m N_m=  \sum_m \sum_n^{\ell_{max}} |\beta_{mn}|^2~.
\eeq
An independent check can be made by confirming the unitarity condition which is satisfied by the Bogolubov coefficients:
\beq
\sum_n^{\ell_{max}} (|\alpha_{mn}|^2 -  |\beta_{mn}|^2) =1\,,
\label{unitarity}
\eeq
where $\alpha_{mn}$ and $\beta_{mn}$ are defined above.

\begin{figure}[t]
\scalebox{0.45}{\includegraphics{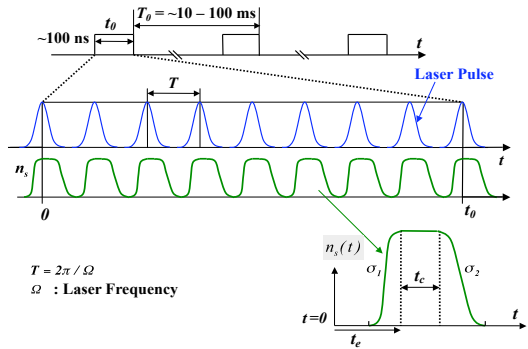}}
\vspace{-1.0cm}
\caption{\label{pulse} (Color online) Example of pulse train (duration $t_0$), the laser pulse (period $T$) and asymmetric surface charge density, $n_s(t)$, which is assumed as a flat-top {plateau} sandwiched by {two} asymmetric half Gaussians.}
\vspace{-0.25cm}
\end{figure}
 
\begin{figure*}[t]
\vspace{-0.5cm}
\scalebox{0.45}{\includegraphics{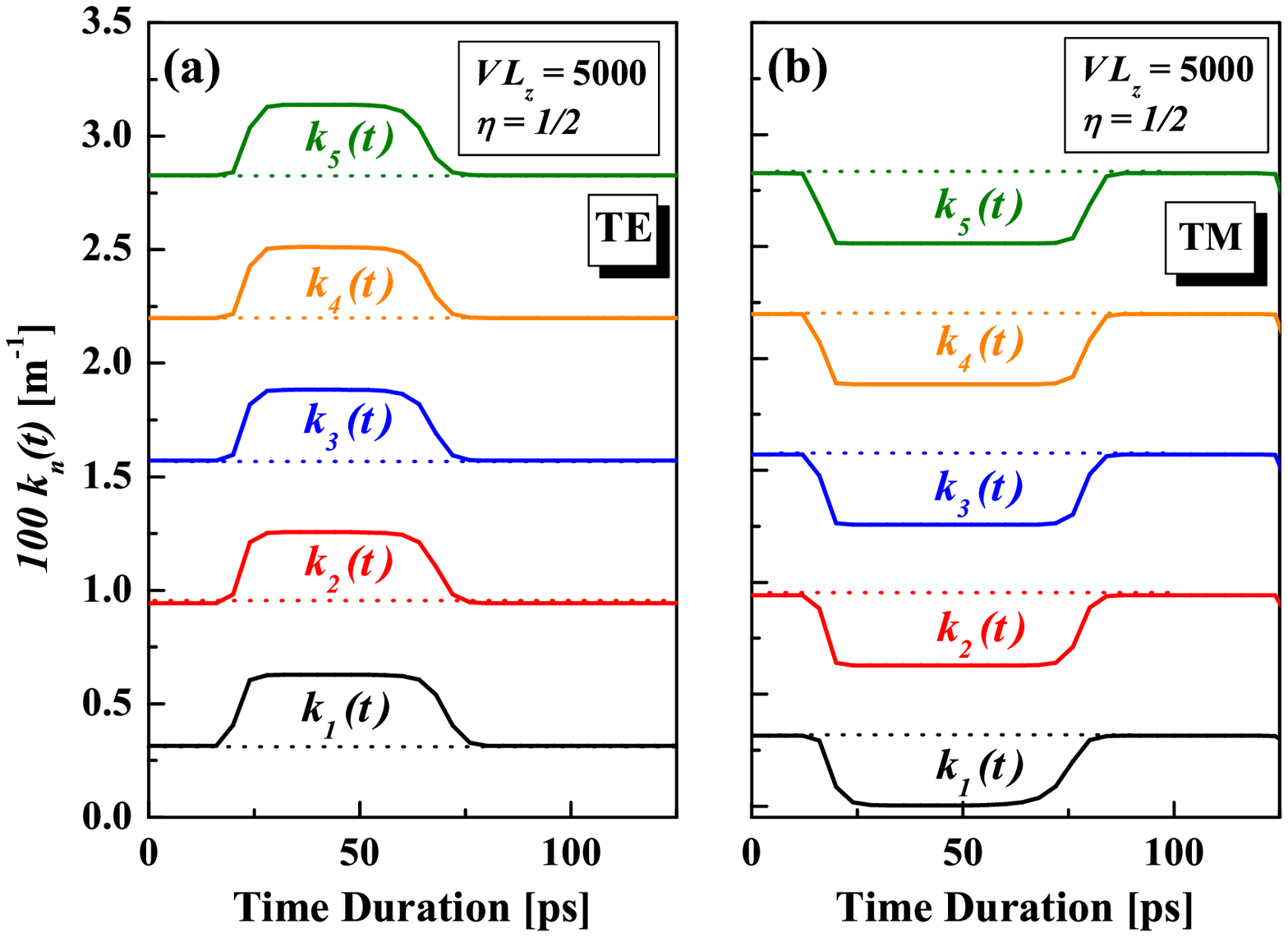}}
\scalebox{0.45}{\includegraphics{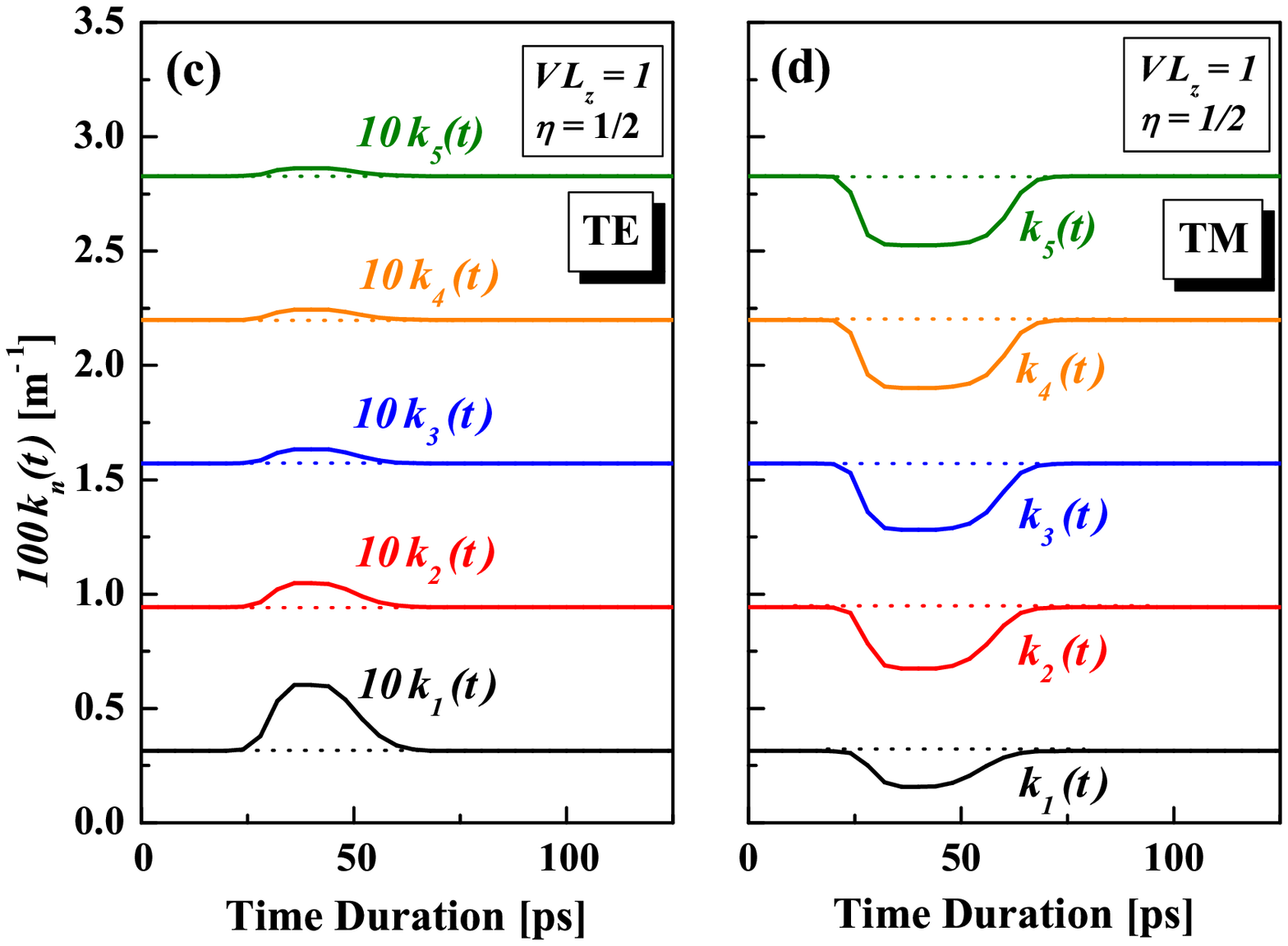}}
\vspace{-0.25cm}
\caption{\label{freq}  (Color online) Frequency variation for various $k_{n}$ for TE and TM modes with two different laser powers, in dimensionless units $V_{\rm max} L_z=5000 $ and $V_{\rm max} L_z=1$. Dashed lines denote baselines corresponding to stationary values. In Fig. \ref{freq} (c) the frequency variation is magnified by a factor of $10$. }
\vspace{-0.25cm}
\end{figure*}
 
\section{Experimental Remarks }
\label{assump}

\par Our primary concern is focused on cavity photons for  TE$_{111}$ and TM$_{111}$ modes, which have the lowest frequency where both TE and TM modes coincide. We shall consider an idealized lossless cavity with a rectangular shape ($L_x = L_y = L_z/2= 0.05$ m), which is bisected by a thin  plasma sheet (SD) placed at $z = d$ from one of the cavity walls. TE$_{111}$ and TM$_{111}$ modes have an angular frequency of $\omega_{111} = 3.05 \times 10^{10} s^{-1}$ implying that we need a laser frequency of around $\Omega=2\omega_{111}  = 6.1\times10^{10} {\rm [s^{-1}]}$ corresponding to a period of $T = 111.1$ [ps] (see Fig. \ref{pulse}). 

\par In the present study, we assumed an infinitesimal thickness for the SD with a time-varying conductivity at a temperature below $1$ K, where all the conduction electrons (those that are excited from the valence band and interact with the cavity photons) are expressed by the surface charge density (areal density), $n_s(t)$ (see Fig. \ref{pulse}). This is estimated assuming a laser wavelength of $860$ nm and a penetration depth $d_s\sim5\, \mu$m for GaAs and is compatible with the diffusion length of the conduction electrons during laser-pulsed irradiation. In practice the thickness of the SD is about $0.5\,$mm and the penetration depth of the $\omega_{111}$ cavity photons in GaAs ({carrier density} $\sim10^{23} {\rm m}^{-3}$) when laser irradiated is estimated to be $\sim 50\, \mu$m for a laser power of $100 \mu$J/pulse. {As discussed later}, the plasma frequency $\omega_p=\sqrt{e^2 n_s/m^* d_s}$ even for a low laser power of $0.01\, \mu$J/pulse is much larger than $\omega_{111}$ (by more than $\sim10$ times) and hence almost all the cavity photons are reflected by the plasma sheet. Thus,  reflection takes place within a depth much less than the penetration depth in the SD when laser-irradiated. This is certainly small enough as compared with the scale of a microwave cavity of length $\sim100$ mm.  Taking into account all these points, the assumption of an infinitesimal thickness for the SD should be a good approximation to the real situation.\footnote{Dodonov and Dodonov developed detailed analyses on the dissipation of cavity photons in a SD in a series of papers \cite{Dodonov:2005ob,Dodonov:2006is,Dodonov:2006jb}. They discussed the temperature rise of a cavity caused by laser irradiation, which depends on the laser power (in their case assumed of order mJ/pulse), thermal conductivity, and size of the SD. However, in our proposed experiment, the laser power will be  constrained to a range of $\mu$J/pulse with the duration of the pulse train set to the order of $100$ ns ($\sim1000$ pulses) which is compatible with the SD cooling time \cite{Dodonov:2005ob} and thus, the temperature rise is likely to be very small.}


\par Another important factor is the relaxation time ($\sigma_2$) to go from metallic to semiconducting states via recombination immediately after pulse irradiation, where in general $n_s(t)$ has an asymmetric profile with a tail (see Fig. \ref{pulse}). {For unitary evolution the need for an asymmetric profile has been stressed by Dodonov \& Dodonov\cite{Dodonov:2005ob,Dodonov:2006is}; however, they} also pointed out \cite{Dodonov:2006jb} significant losses of photons in a semiconductor slab due to an  asymmetric $n_s(t)$ profile with a long tail and thus, the need for a short relaxation time of less than about $20$ ps to detect DCE photons with a frequency of $2.5$ GHz. Recently, Agnesi et 
al. \cite{Agnesi:2008ja} reported that the relaxation time could be reduced to $\sim 20$ ps for GaAs irradiated by neutrons. Alternatively, the relaxation time can be shortened by making a deep (trapping) level in the band gap of a semiconductor, which is introduced by high energy electron irradiation and doping with Au etc. Indeed, it was reported that GaAs irradiated with $200$ MeV Au${}^{+}$ {ions gives a} relaxation time shorter than $4$ ps at $7$K \cite{mangeney:4711}. Thus it is possible to make a profile without a long tail for  
$V(t)L_z= e^2 n_s(t) L_z/ m^*$ which oscillates between $0$ and  $V_{\rm max} L_z$ ranging typically from $1$ to $10,000$ in our simulations. {For the $n_s(t)$ profile we have assumed a flat-top {plateau} sandwiched by two asymmetric half Gaussians, as {depicted} in Fig. \ref{pulse}, with an off-set of $t_e = 35$ ps, plateau of $t_c=7$ ps and standard deviations of $\sigma_1=4$ and $\sigma_2=8$ ps, respectively. }

\par Much of the motivation for this work is based on a novel DCE detection system, using highly-excited Rydberg atom (Rb or K) beams working as a single photon detector, which has already been utilized to explore the dark matter axion \cite{Bradley:2003kg}. The cavity is cooled down to $\sim100$ mK by a dilute refrigerator using ${}^3$He combined with a pulse tube to suppress the number of thermal blackbody photons to be less than $0.5$. Note that as well as DCE photons, thermal blackbody photons will also be enhanced parametrically \cite{Plunien:1999ba}, so there is no {\it a priori} reason to do experiments at low temperatures. Our interest; however, is to detect the DCE as a pure vacuum fluctuation effect. At the same time, a low laser power is inevitable in order to keep the cavity at a low temperature and to achieve a high quality factor $Q$ of better than $10^6$, which guarantees a lifetime of longer than a few milliseconds for the resonant microwave photons created.

\par A high-sensitivity photon detector is required to observe actual DCE photons and a highly excited Rydberg-atom beam is one of the most promising techniques, by working as a single photon detector in the microwave regime. Our concern in this work is to find the optimum conditions for parametric enhancement of DCE photon creation by numerical analysis and thus, we evaluate the number of photons created dependent on SD position, laser power, and the range of driving laser frequency giving DCE photon creation.  

\par In general we must find solutions to $k_{n}(t)$ numerically in Eq. (\ref{transTETM}); however, in the limit $\eta\ll 1$ (or for low laser power) it is possible to find a linearized solution for TE modes only. This is similar to the approach of Crocce et al. \cite{Crocce:2004jq} who for $\eta=1/2$ with ${e^2 n_s(t)/ m^*}$ ranging from $10^{10}-10^{16}$ m${}^{-1}$, which is valid for doped semiconductors at room temperature, also found a linearized solution. Under practical experimental
conditions; however, as mentioned, we require a low temperature of around $100$ mK to suppress thermal background radiation and to get a high $Q$ value. If a laser power of $100~\mu$J/pulse is also assumed, then the actual range is estimated to be from $0$ (non laser-irradiated) to around $10^5$ m$^{-1}$ and hence, a perturbative analysis is not possible (unless $\eta \ll 1$ for TE modes). Indeed, this was part of our motivation for carrying out a full numerical and non-perturbative analysis.

\par It is also important to estimate numerically the frequency shift, which was recently discussed in \cite{Dodonov:2006jb,Crocce:2004jq, Yamamoto:2008}. For the mechanical vibrations of a cavity wall the average frequency $\tilde \omega_n = \tilde k_n c$  is close to that  of the stationary angular frequency, $\omega_n^0$, see Eq.\,(\ref{statfreq}). However, for an SD irradiated by a pulsed laser this is not the case, because the time-variation of $k_n ( t )$  shifts from the baseline frequency asymmetrically (dashed lines in Fig. \ref{freq}) with a positive shift for TE$_{11n}$ and a negative shift for TM$_{11n}$ modes, see Fig. \ref{freq} for the $\eta=1/2$ case. Hereafter, we denote $\tilde \omega_n - \omega_n^0$ the ``frequency shift," which should be taken into account when tuning the driving period of the pulsed laser for parametric photon creation. 

\begin{figure*}[ht]
\scalebox{0.475}{\includegraphics{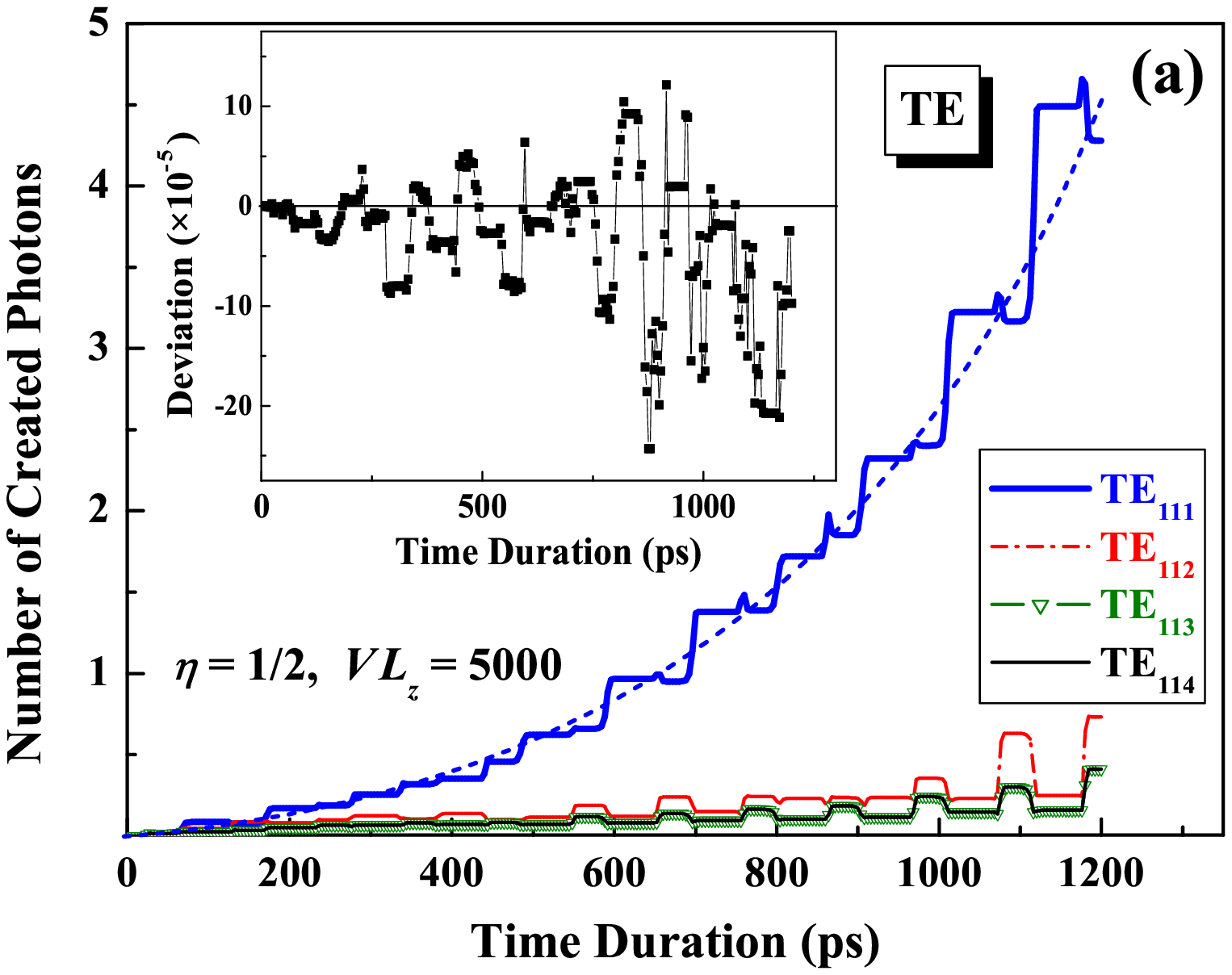}}
\scalebox{0.475}{\includegraphics{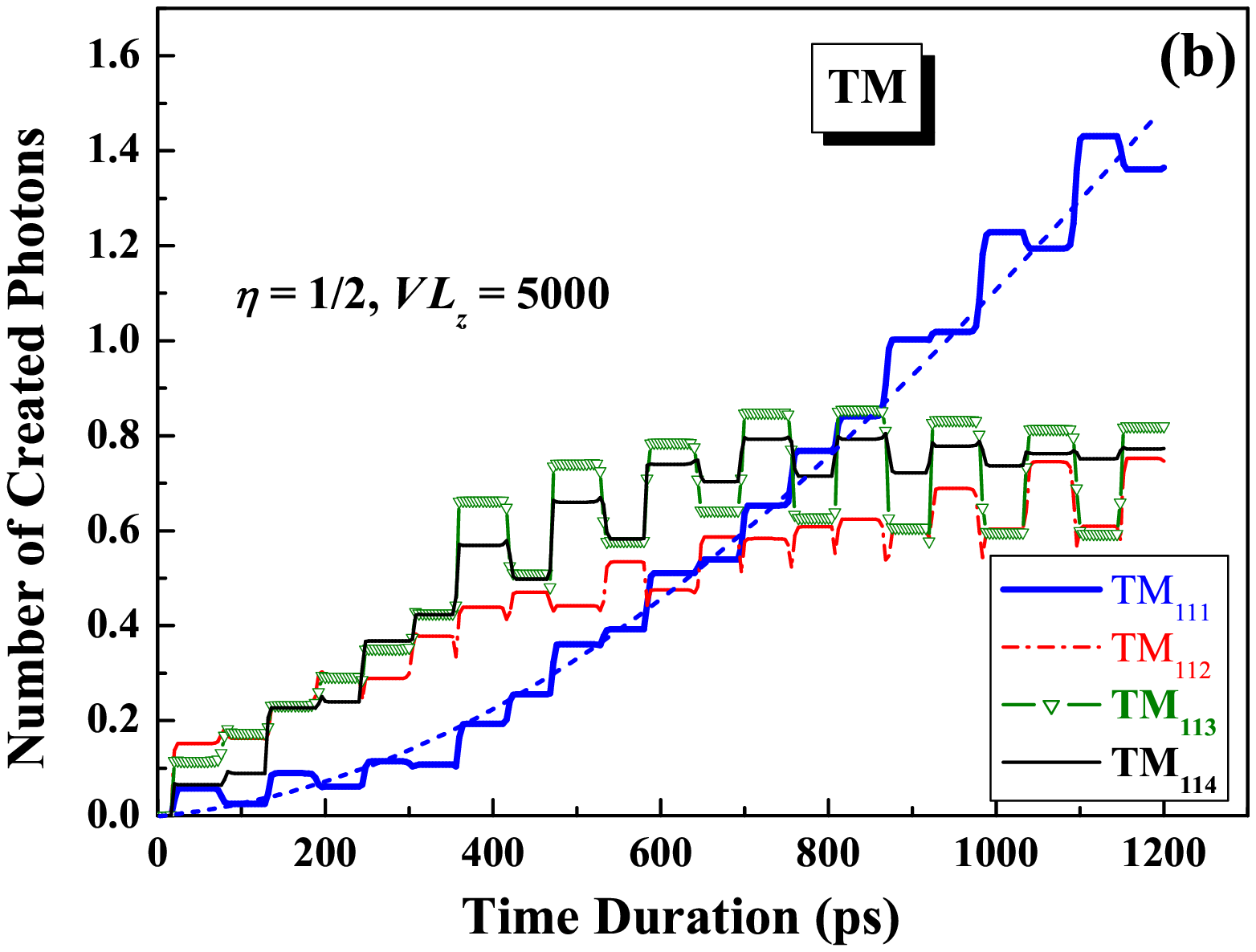}}
 \vspace{-0.5cm}
\caption{\label{timeplot} (Color online) $N_{111}(t)$ with  $\eta=1/2$ and $V_{\rm max}L_z = 5000$ calculated for (a) TE$_{111}$ cavity mode (following the dashed line) with higher modes giving lesser contributions, respectively; and (b) TM$_{111}$ cavity mode (following the dashed line), where TM$_{112}<  {\rm TM}_{114}<{\rm TM}_{113}$ at $500$ ps, for example.  Fig. 3 (a) inlay: Unitarity constraint, $d_m=1- \sum_n^{\ell_{max}} (|\alpha_{mn}|^2 - |\beta_{mn}|^2)$ in this case with $m=1$ and cutoff $\ell_{max} = 51$. For TM$_{111}$ the unitarity constraint is within $2\times10^{-4}$ (not shown here). }
\vspace{-0.5cm}
\end{figure*}
\begin{figure*}[ht]
\scalebox{0.45}{\includegraphics{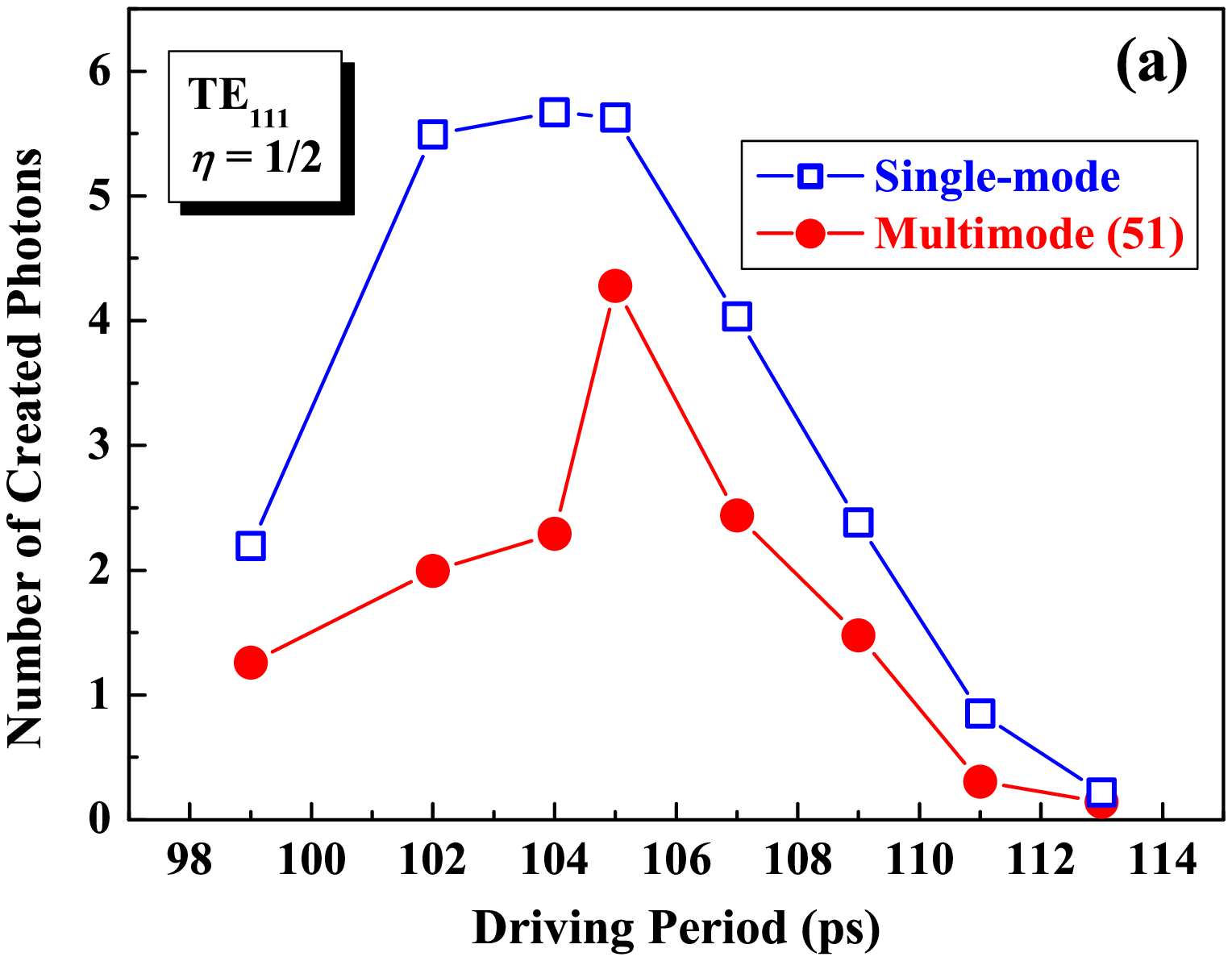}}
\scalebox{0.45}{\includegraphics{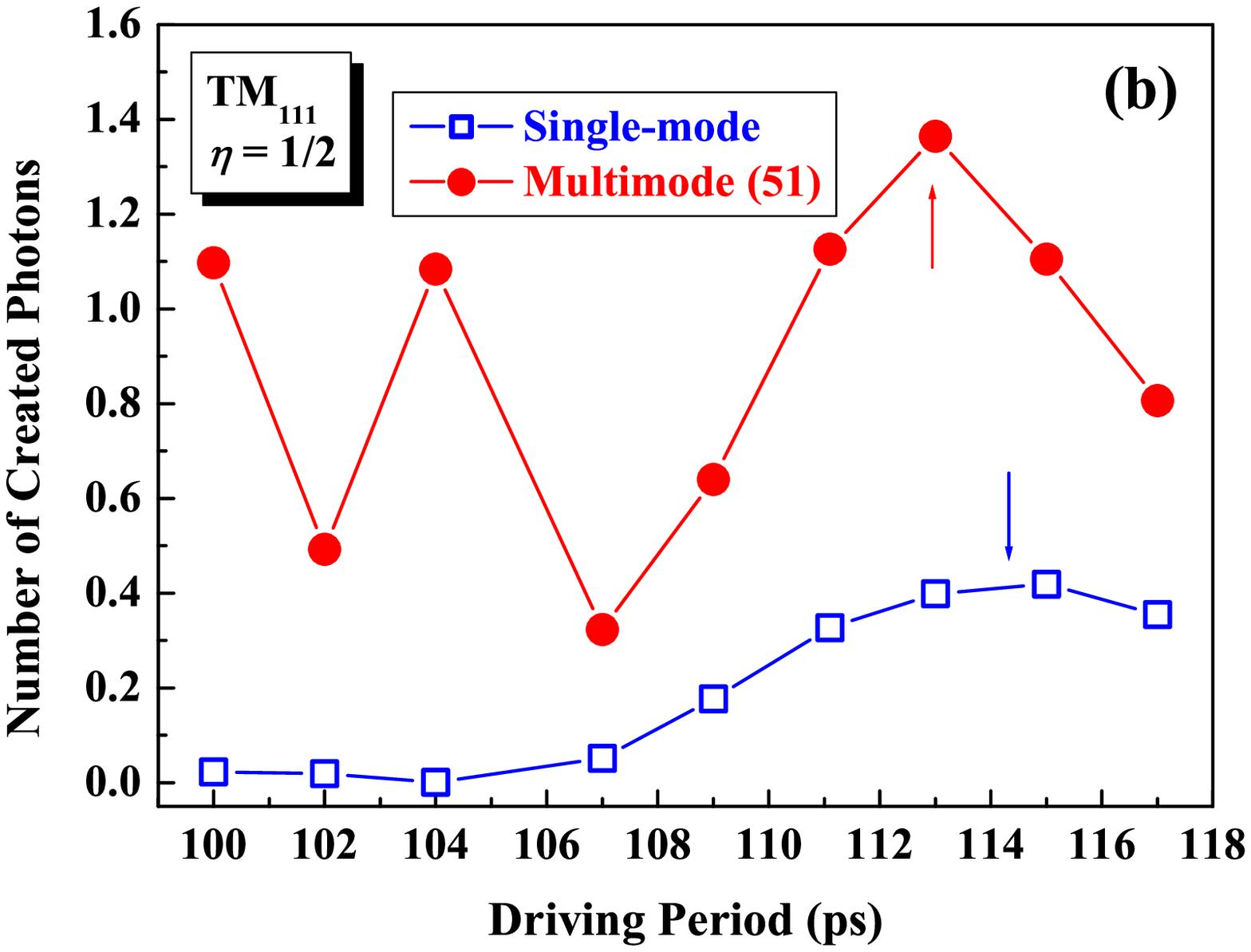}}
 \vspace{-0.5cm}
\caption{\label{tuning} (Color online) $N_{111}(t\sim 11T)$ for (a) TE$_{111}$ and (b) TM$_{111}$ modes at $\eta=1/2$ with $V_{\rm max}L_z = 5000$. Open squares and filled circles denote single mode and multimopde ($\ell_{max} = 51$) couplings, respectively, and  arrows indicate optimum conditions. }
\vspace{-0.5cm}
\end{figure*}
\begin{figure}[h]
\scalebox{0.47}{\includegraphics{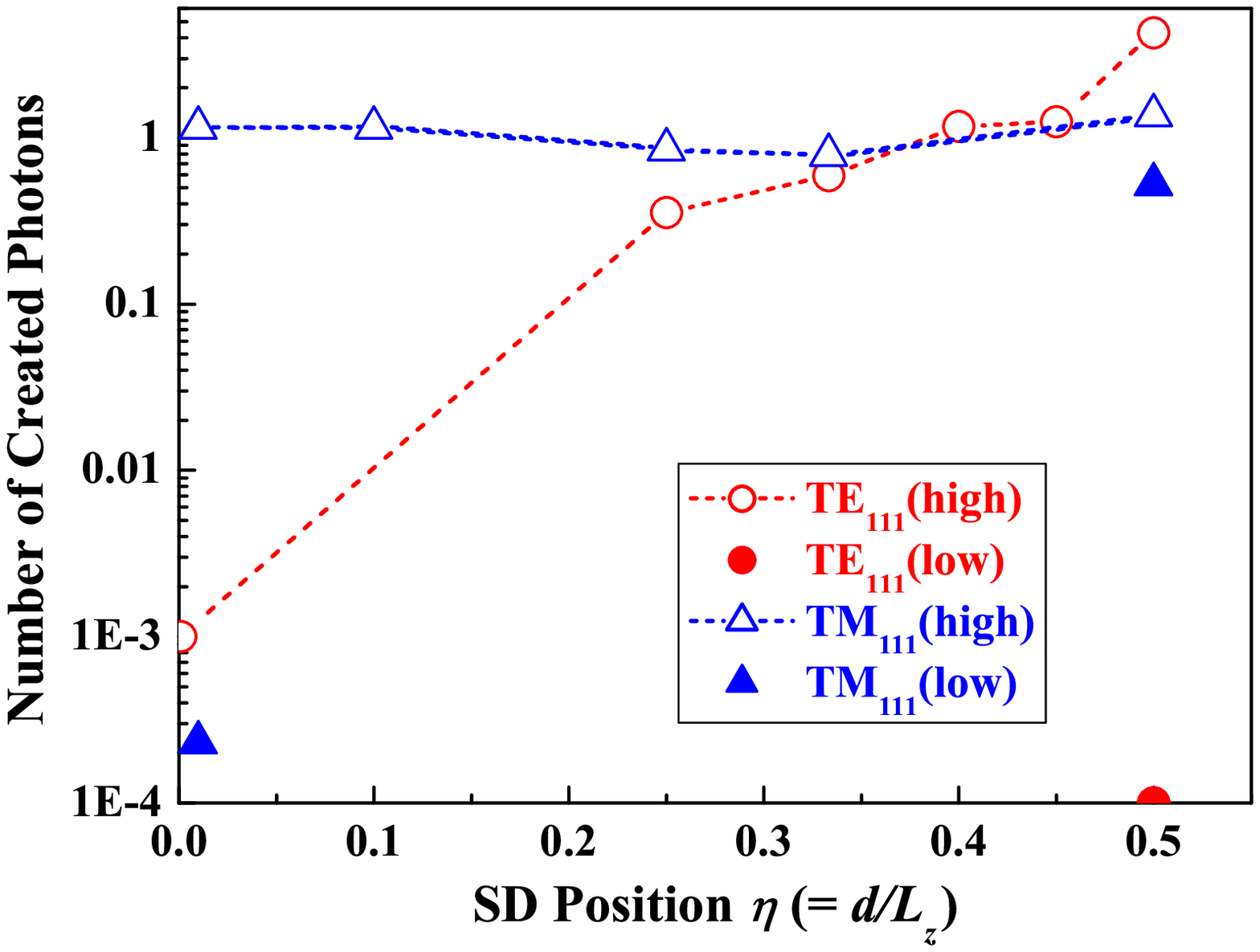}}
\vspace{-0.75cm}
\caption{\label{place} (Color online) $N_{111}(t\sim 11T)$ for TE$_{111}$ and TM$_{111}$ modes dependent on SD location for a given optimum laser period $T$ with $V_{\rm max}L_z = 5000$. Filled triangles and circles denote a low laser power with $V_{\rm max}L_z = 1$. }
\vspace{-0.35cm}
\end{figure}

\section{Results \& Discussion}
\label{analysis}

\par The time variation of the eigenvalue $k_{n}(t)$ for both TE$_{11n}$ and TM$_{11n}$ modes is shown in Fig. \ref{freq}, which reflects the interaction between the cavity photons and the conduction electrons excited by the pulsed laser.  Figs. \ref{freq} (a) and (b) show  $k_{n} (n = 1, 2, 3, 4, 5)$ for TE and TM modes, respectively, assuming  $\eta=1/2$ and a $V_{\rm max} L_z$  value of $5,000$ ($50 \mu$J/pulse).  Here, we should note that the plasma frequency   $\omega_p \simeq 3\times 10^{13}$ s$^{-1}$ is much larger than the frequency of the cavity photons of interest
and thus, a pulsed laser will act as a high-frequency switching (transmission/reflection) actuator. Laser irradiation leads to a positive change in  $k_{n}$ for TE$_{11n}$, while negative changes occur for TM$_{11n}$ modes from each baseline, which corresponds to the stationary $k_n^0$ value for the cavity photons.  The width and variation of $k_{n}(t)$ are almost independent of $n$ for both TE and TM modes, but they gradually get smaller with increasing $n$.  However, for TM modes the variation  is significantly larger than that for TE modes, which is due to the strong coupling of the TM electric field component, $E_z$, to the conduction electrons in the plasma.  

\par From an experimental viewpoint, it is desirable to decrease the laser power for the suppression of {temperature rises} in the cavity. Shown in Figs. \ref{freq} (c) and (d), respectively, are the variation of TE$_{11n}$ and TM$_{11n}$ modes, calculated assuming a very low laser power of $0.01\mu$J/pulse ($V_{\rm max} L_z =1$) at $\eta=1/2$.  Also in this case, the plasma frequency of  $\omega_p \simeq 4\times 10^{11}$ s$^{-1}$  is still much larger than the frequency of TE$_{11n}$ and TM$_{11n}$ modes.  As expected, the width and frequency variation of $k_n c$ for TE$_{11n}$ modes drastically decreases.  Interestingly; however, such an extreme reduction of laser power does not decrease the frequency variation for TM$_{11n}$ modes, although the widths are somewhat reduced.  {Such a large frequency variation leads to a large enhancement of photon creation and is intimately related to the TM plasma sheet boundary conditions.} Indeed, as discussed later, the number of created photons for TM$_{111}$ does not decrease even with a  reduction in laser power of order $1/5000$ ($0.01\mu$J/pulse), which is quite different from the TE$_{111}$ case. 


\par The time evolution of the number of photons for TE$_{111}$/TM$_{111}$ modes with SD position at the midpoint, $\eta = 1/ 2$, is depicted in Figs. \ref{timeplot} (a) and (b) respectively, up to a time lapse of $\sim11 T$ with a driving period of $T=105$ ps (frequency of $9.52$ GHz) for TE$_{111}$ and  $T=113$ ps (frequency of $8.85$ GHz) for TM$_{111}$ modes. Here, the stationary $(1, 1, 1)$ mode in the cavity ($0.05\times0.05\times0.10 \,{\rm m}^3$) corresponds to a frequency of 4.5 GHz and thus, the driving laser frequency is expected to be around $9.0$ GHz (with period $T = 111.11$ ps). These optimum driving periods are correlated with shifted frequency {$\tilde k_n c$} corresponding to half of the driving periods of $104.2$ and $116.1$ ps for TE$_{111}$ and TM$_{111}$ {modes}, respectively, as can be derived from Figs. \ref{freq} (a) and (b). As we mentioned previously such a frequency shift comes  from the interaction between the cavity photons and conduction electrons. Other mode couplings ($112, 113$, and $114$) shown in 
Fig. \ref{timeplot} (a) are also slightly enhanced with time.  
Some intermode couplings satisfying  $\omega_{111}+\omega_{11n} = j \Omega_{111} $ ($j$: integer) enhance photon creation, while those far from the resonant condition damp the number of photons created. 
An important check of the numerical results is the unitarity constraint, Eq. (\ref{unitarity}), which leads to an independent check of the convergence of the cut-off, $\ell_{max}=51$, in the mode sums. The inlay in Fig. {\ref{timeplot} (a) shows the deviation from the unitarity condition, showing sufficient convergence (deviation: $ \leq 2\times 10^{-4}$). Similar convergence is also confirmed for TM$_{111}$ modes (not shown here).

\par The number of created photons {$N_{111}(t)$ for $t \sim11 T$} dependent on driving laser period, $T$ 
 is shown in Fig. \ref{tuning} for both TE$_{111}$/TM$_{111}$ modes at $\eta = 1 / 2$ (midpoint). 
Interestingly, multimode coupling suppresses photon creation for TE$_{111}$ modes, while it enhances the number of photons for TM$_{111}$ modes.  
Such an enhancement is caused by constructive ($\omega_{111}+\omega_{11n}\cong j \Omega_{111}$) mode couplings, as mentioned before and shown in Figs. \ref{timeplot} (a) and (b). {It should be emphasized that the optimum driving period, which is correlated with the frequency variation and intermode couplings, is different from the resonant one corresponding to the unperturbed field eigenfrequency (stationary mode frequency) for all values of $\eta$.}  The number of photons for the TE$_{111}$ mode takes a maximum value of $4.3$ at $T \sim105$ ps, which is almost 15 times that at $111.1$ ps ($\Omega=2\omega_{111}$). In contrast, N$_{111}$ for the TM case oscillates periodically and is significantly smaller than the maximum of N$_{111}$ for the TE case (at $\eta=1/2$), because the higher modes such as $k_2,~k_3,...$ still have large frequency variation for TM modes, see Fig.~\ref{freq}.  Importantly, however, the driving frequency of  $\Omega=2 \omega_{111}$ ($111.1$ ps) still gives a large number of photon creation for TM$_{111}$ modes. Such a  behavior is probably due to the different natures of TE modes (with $B_z$ component) and TM modes (with $E_z$ component) that satisfy different {\it jump} conditions.
The present results clearly show that parametric photon creation takes place for TE$_{111}$ and TM$_{111}$ at $\eta=1/2$  for a relatively wide range of driving periods.  This is quite advantageous for the experimental detection scheme of DCE photons using a plasma sheet irradiated by a pulsed laser.

\par The parametric photon creation rate depends on the location of the SD, $\eta=d/L_z$, as shown in Fig. \ref{place} for TE$_{111}$ and TM$_{111}$ modes, respectively. Here, the driving period $T$ is chosen to give the optimum value. The $N_{111}$ value for TE$_{111}$ modes appears to increase drastically with increasing $\eta$ and reaches a maximum at $\eta = 1/2$ (the midpoint), while a weak dependence on SD position is seen for TM$_{111}$ modes. We should note that $N_{111}$ at the midpoint for TE$_{111}$ is much larger (by an order of three) than that at $\eta \ll1$ (near the cavity wall). This probably comes from the fact that the TE$_{111}$ mode has a belly (anti-node) at the midpoint  ($\eta = 1 / 2$).  For $\eta \ll 1$, $N_{111}(t)$ for TM$_{111}$ is much larger than that for TE$_{111}$ modes, as predicted thus far \cite{PhysRevA.47.4422,PhysRevLett.93.193601,Dodonov:2005ob}. As we previously mentioned, a decrease in laser power does not reduce the amplitude of $k_{n}$  for TM$_{11n}$ modes, although it does for the TE case.  Indeed, a decrease in laser power of the order of 1/5000 leads to a reduction of only half the number of photons for the TM$_{111}$ mode at the midpoint, as shown in Fig. \ref{place} (filled triangle). Unfortunately; however, {a very low laser power reduces almost proportionately} the number of N$_{111}$ photons for TE$_{111}$ at  $\eta=1/2$ and for TM$_{111}$ with $\eta \ll 1$, as indicated in Fig. 5.  As mentioned above, our numerical calculations without any perturbations show that the best location for the SD is the midpoint of the cavity for both TE$_{111}$ and TM$_{111}$, in particular for TM$_{111}$ mode{s} even with a low laser power. 

\section{Conclusion}
\label{conc}

\par In conclusion our numerical results, without any perturbations, indicate that the midpoint is the best location for a semiconductor diaphragm (SD) for {both} TE$_{111}$ and TM$_{111}$ modes, particularly for TM modes even with a low laser power of around $0.01~\mu$J/pulse.   It is emphasized that such a low laser power suppresses heat losses and makes it possible to keep the cavity at a low temperature. The location of the midpoint for the SD is an advantageous configuration to suppress background signals, because such a geometry enables us to separate the cavity into two parts, one of which is chosen as a photon-creation/detection chamber {with the other used} as a laser irradiation space. 

\par {Another important} point highlighted here are the frequency shifts mainly caused by the time-varying areal density of conduction electrons, $n_s(t)$, which are excited by the pulsed laser. This requires careful tuning of the driving laser frequency, although we do find a relatively wide range of driving period allowed to induce parametric photon creation for both TE$_{111}$ and TM$_{111}$ modes.  The present numerical analysis clearly shows that an SD irradiated by a pulsed laser behaves as a high-frequency switching (transmission/reflection) actuator \cite{Sturge:1962,braggio:4967} instead of a mechanically vibrating cavity wall (mechanical oscillations with frequencies of GHz order seem technically difficult to achieve). Also see \cite{Segev:2007,saito:033605} for other alternate experimental proposals.

\par We have designed a new DCE detection system using highly-excited Rydberg atom (Rb or K) beams, acting as a single photon detector with a high sensitivity and a cavity made of niobium, which can be cooled down to $\sim100$ mK {using} a dilute refrigerator combined with a pulse tube 
 and thus, we expect a high $Q$ value better than $10^6$. (Note, $Q$-factors as high as $3\times10^{10}$ have been reported in superconducting cavities \cite{Rempe:1990, Brune:1996} leading to a lifetime of order $\sim1$ s, but the insertion of the SD will reduce this value.) The experimental set-up is now {in progress}. 

\acknowledgements
\par The authors acknowledge valuable discussions with K.~Yamamoto  concerning frequency shifts. This work is supported by Grant-in-aid for Scientific Research (B) by the Ministry of Education, Japan.

\appendix
\section{Plasma Sheet Model}
\label{AppA}

\par In this appendix we briefly discuss the plasma sheet model and derive the {\it jump} conditions for the TE and TM fields ($\Psi,\Phi$) given in Eqs. (\ref{TEbc}) and (\ref{TMbc}). 
The two-dimensional ($(x, y)$ plane) current density $\bb K$ for conduction electrons {with relaxation time, $\tau$, under the presence of an external electric field $\bb E$ can be expressed using the vector potential $\bb A$ \cite{namias:898}:}
\beq
{\bb K} 
= {e^2 \tau n \over m^*} {\bb E_\bot} 
= {e^2 n_s\over m^*} \mathbf A_\bot ~.
\label{curr}
\eeq
Combining the continuity condition $\dot \sigma+ \bb\nabla \cdot \bb K=0$ with the Lorenz gauge {condition} then leads to 
\beq
\dot \sigma 
= -{e^2 n_s\over m^*} \bm\nabla\cdot {\bb A_{\bot}}
={e^2 n_s\over m^*} \partial_t A_0
~.
\eeq
where $n$ is the bulk charge density and  $A_0$ is the scalar potential. As a result, the surface charge density, $\sigma$, is given by
\beq
\sigma = {e^2 n_s\over m^*} 
A_0~.
\label{surf}
\eeq
\par It is convenient to now define two Hertz vectors $\mathbf\Pi_e$ and $\mathbf\Pi_m$ as \cite{Crocce:2005htz,Nisbet:1955}:
\beq
A_0=-{1\over \epsilon} \bm\nabla\cdot \mathbf \Pi_e\,, \quad\quad
{\mathbf A}=\mu{\partial \mathbf \Pi_e \over \partial t} + \bnab\times {\mathbf \Pi_m}~.
\label{gaugeflds}
\eeq
where $\veps$ and $\mu$ are the {bulk} electric permittivity and magnetic permeability of the SD, respectively. For electromagnetic waves propagating along the $z$-axis of a rectangular cavity, the Hertz vector potentials can be reduced to the following 1D expression 
\bea
{\mathbf \Pi}_e=\Phi ~\hat {\bb e}_z\qquad\qquad\qquad {\mathbf \Pi}_m=\Psi ~\hat {\bb e}_z~,
\label{Hvect}
\eea
where $\hat \bb e_z$  is a unit vector in the $z$-direction. If $\Psi~(\bb \Pi_e)$ and $\Phi~(\bb \Pi_m)$ are given the fields $\mathbf E$ and $\mathbf B$ are expressed as 
\begin{widetext}
\bea
{\mathbf E}&=&-\nabla A_0-\partial_t {\bb A}= \Biglb[\frac 1 \veps \partial_x\partial_z \Phi- \partial_y\partial_t\Psi\Bigrb] \unit e_x +\Biglb[\frac 1 \veps \partial_y\partial_z \Phi+ \partial_x\partial_t\Psi \Bigrb]\unit e_y 
- \frac 1 \veps \Biglb[ {\partial^2\over\partial x^2} +{\partial^2\over\partial y^2} \Bigrb]\Phi\, \unit e_z
\nn
{\mathbf B} &=&  \nabla\times {\bb A} =
\Biglb[\partial_x\partial_z\Psi+\frac 1 \mu\partial_y\partial_t \Phi\Bigrb] \unit e_x  + 
 \Biglb[\partial_y\partial_z \Psi-\frac 1 \mu\partial_x\partial_t \Phi\Bigrb] \unit e_y  - 
\Biglb[ {\partial^2\over\partial x^2} +{\partial^2\over\partial y^2} \Bigrb]\Psi\, \unit e_z ~,
\label{Zfields}
\eea
\end{widetext}
where we have used the fact that
$
\bnab_{\bot}^2 \Phi (\Psi)= \Biglb[ {\partial^2\over\partial x^2} +{\partial^2\over\partial y^2} \Bigrb] \Psi (\Phi)=-\bb k^2_\bot \Psi (\Phi).
$ 
Note that $\Phi$ and $\Psi$ satisfy the classical 3D wave equations $\biglb[\nabla^2 - \veps\mu \,\partial_t^2 \bigrb] \Psi (\Phi)=0$. The boundary conditions for a stationary interface are, e.g., see \cite{namias:898}:
\bea
( {\mathbf D}_{2}-{\mathbf D}_1)\cdot {\unit e}_z&=&\sigma ,
\quad\quad
\qquad( {\mathbf B}_{2}-{\mathbf B}_1)\cdot {\unit e}_z=0~,
\nn
{\unit e}_z\times( {\mathbf H}_{2}-{\mathbf H}_1)
&=&{\mathbf K},
\quad\quad
{\unit e}_z\times( {\mathbf E}_{2}-{\mathbf E}_1)
 =0~.\nn
\label{BCs}
\eea
{where ${\bb E}$ and ${\bb H}$ are the electric and magnetic fields; and  ${\bb D}$ and $\bb B$ are the electric and magnetic flux densities, respectively.} Substituting in our values of $\sigma$ and ${\bb K}$ we find the following boundary interface conditions {at $z=d$},
\bea
&&{\rm disc}~ \Phi(d) =
-\mu {e^2 n_s\over {\bb k}_\bot^2 m^*} \partial_z \Phi  |_{z=d} \cong 
-\mu_0 {e^2 n_s\over {\bb k}_\bot^2 m^*} \partial_z \Phi |_{z=d} \nn
&&{\rm disc}~ B_z = {\bb k}_\bot^2{\rm disc}~ \Psi(d)=0
\\
&& {\rm disc}~ \partial_z \Psi |_{z=d} = 
\mu {e^2 n_s\over m^*} \Psi(d) \cong 
\mu_0 {e^2 n_s\over m^*}  \Psi(d) 
\nn
&& {\rm disc} ~\partial_z \Phi |_{z=d}=0
\nonumber
\eea
The above boundary conditions come from the following differential equations ($\mu_0 \to 1$)
\bea
 \bnab_{\bot}^2\Psi+ \partial_z^2\Psi-\partial_t^2\Psi&=&{e^2 n_s\over m^*} \delta (z-d)\Psi(d)~,\nn
 \bnab_{\bot}^2\Phi+  \partial_z^2\Phi-\partial_t^2\Phi&=&{1\over {\bf k}_\bot^2}{e^2 n_s\over m^*}\delta'(z-d)\Phi(d)~,\nn
 \label{TETMode}
\eea
{which are those presented in Eqs. (\ref{ode}) and (\ref{TMode}) for TE and TM modes, respectively.}

\newpage
\bibliography{SDPlasma}{}

\end{document}